%% file: spie_hires_v2.tex
\documentclass[a4paper]{spie}
\usepackage{amsmath}
\usepackage[]{graphicx}
\usepackage{mdwlist} % Serve per avere {itemize*} che fa spazi minori
\title{Trade-off study for high resolution spectroscopy in 
the near infrared with ELT telescopes: seeing-limited vs. diffraction 
limited instruments}

\author{
 N.~Sanna\supit{a},
 E.~Oliva\supit{a},
 F.~Massi\supit{a},
 G.~Cresci\supit{a},
 L.~Origlia\supit{b}
\skiplinehalf
\supit{a} INAF -- Osservatorio di Arcetri, 
   Largo E. Fermi 5, I-50125 Firenze, Italy;\\
\supit{b} INAF -- Osservatorio di Bologna, 
  Via Ranzani 1, I-40127 Bologna, Italy\\
}

\authorinfo{contact e-mail of author: sanna@arcetri.inaf.it}

\begin{document}
  \maketitle

%%%%%%%%%%%%%%%%%%%%%%%%%%%%%%%%%%%%%%%%%%%%%%%%%%%%%%%%%%%%%

\begin{abstract}
HIRES, a high resolution spectrometer, is one of the first five instruments foreseen
in the ESO roadmap for the E-ELT.
This spectrograph should ideally provide full spectral coverage from the UV limit to
2.5 microns, with a resolving power from R$\sim$10,000 to R$\sim$100,000. 
At visual/blue wavelengths,
where the adaptive optics (AO) cannot provide an efficient light-concentration,
HIRES will necessarily be a bulky, seeing-limited instrument.
The fundamental question, which we address in this paper, is whether the same
approach should be adopted in the near-infrared range, or HIRES should only
be equipped with compact infrared module(s) with a much smaller aperture, taking 
advantage of an AO-correction.
The main drawbacks of a seeing-limited instrument at all wavelengths are:
\textit{i)} Lower sensitivities at wavelengths dominated by thermal background (red part of
the K-band).
\textit{ii)} Much higher volumes and costs for the IR spectrograph module(s).
The main drawbacks of using smaller, AO-fed IR module(s) are:
\textit{i)} Performances rapidly degrading towards shorter wavelengths (especially J e Y
bands).
\textit{ii)} Different spatial sampling of extended objects (the optical module see a much larger
area on the sky).
In this paper we perform a trade-off analysis and quantify the various effects that
contribute to improve or deteriorate the signal to noise ratio. In particular, we
evaluate the position of the cross-over wavelength at which AO-fed instruments starts
to outperform seeing-limited instruments. This parameter is of paramount importance for
the design of the part of HIRES covering the K-band.
\end{abstract}

\keywords{Ground based infrared instrumentation, infrared spectrometers,
limiting sensitivities of high resolution infrared instruments for ELTs}

\ \\
\section{INTRODUCTION}
HIRES is one of the first five instruments foreseen in the ESO roadmap for the Extremely Large Telescope
(E-ELT).
This high resolution (HR) spectrometer should ideally provide full spectral coverage from
0.37 to 2.5 microns, with a resolving power ($R=\lambda/\Delta\lambda$) of about 100,000.
It is also expected to work in mid resolution (MR) mode, with $R\sim14,000$. 
A system of fibers optimized for different wavelengths, should feed simultaneously 
(through dichroics) different and independent spectrometers: the UB+V and R+I bands and (cryogenic) 
the Y+J+H and K bands.  

Exoplanets provide one of the outstanding key science cases for HIRES.
The focus will be on characterizing exo-planet atmospheres over a wide range of masses, 
from Neptune-like down to Earth-like (including those in the habitable zones) in terms of chemical 
composition, stratification and weather. The ultimate goal is the detection of signatures of life. 
The extremely high signal-to-noise required to detect the exo-planet atmospheric
signatures has paradoxically pushed this area into the $"$photon-starved$"$ regime with current 
facilities, making the collecting area of the E-ELT essential for achieving the ambitious goals.

A dedicated exposure time calculator (ETC) for the instrument was developed and is described here.
It is a tool to predict the performances of the spectrometer for different 
instrument parameters and environmental conditions.\\
We report here the estimated limiting AB magnitudes in various bands for
different S/N levels both in the seeing-limited and AO cases.

\section{\bf Description of the ETC}

The ETC computes the limiting magnitude achievable at a given wavelength, in 
a given exposure time and at a given signal to noise ratio. The computed
limiting magnitude is in AB units, i.e.
$$ m(AB)= -2.5 
   \log_{10}\left(f_\nu\over {\rm 1\ erg\ cm^{-2}\ s^{-1}\ Hz^{-1}} \right)
     - 48.60 $$
$$ m(AB)= -2.5 
  \log_{10}\left(f_\lambda\over 
      {\rm 1\ erg\ cm^{-2}\ s^{-1}\ \mu m^{-1}} \right)
  - 5.0 \log_{10}\left(\lambda\over {\rm 1\ \mu m}\right) - 12.40 $$
$$ m(AB)= -2.5 
  \log_{10}\left(N_\lambda\over {\rm 1\ photon\ cm^{-2}\ s^{-1}\ \mu m^{-1}} 
    \right)
  - 2.5 \log_{10}\left(\lambda\over {\rm 1 \mu m}\right) + 16.85 $$

The ETC accounts for a number of instrumental and environmental
parameters, such as resolving power, efficiency, pixel sampling, detector 
noise, sky background. These parameters can be modified, thus allowing the
user to quantify their effect on the overall performances of the spectrograph.

The input parameters to the ETC are as follows.

\newcommand{\WL}{{\sl WL}}
\newcommand{\SN}{{\sl SN}}
\newcommand{\EXPTIME}{{\sl EXPTIME}}
\newcommand{\RPOW}{{\sl RPOW}}
\newcommand{\EFF}{{\sl EFF}}
\newcommand{\SKYBCK}{{\sl SKYBCK}}
\newcommand{\TBCK}{{\sl TBCK}}
\newcommand{\EBCK}{{\sl EBCK}}
\newcommand{\DTEL}{{\sl DTEL}}
\newcommand{\COBS}{{\sl COBS}}
\newcommand{\DPIX}{{\sl DPIX}}
\newcommand{\FXCAM}{{\sl FXCAM}}
\newcommand{\FYCAM}{{\sl FYCAM}}
\newcommand{\NDIT}{{\sl NDIT}}
\newcommand{\RON}{{\sl RON}}
\newcommand{\DARKCUR}{{\sl DARKCUR}}
\newcommand{\SLE}{{\sl SLE}}
\newcommand{\DAPE}{{\sl DAPE}}

\begin{itemize}

\item \WL : required wavelength, in $\mu$m 

\item \SN : required signal to noise ratio. This quantity is computed per 
resolution element on the extracted spectrum.

\item \EXPTIME : required exposure time, in hours. This is the total exposure
time. If the observation consists of several, shorter exposures, the user
must update the parameter NDIT. For an IR detector the exposure time for a single
observation ($NDIT=1$) is typically shorter than $\sim$ 30 minutes.

\item \NDIT : number of separate read-outs to achieve the requested exposure 
time. 
It is suggested that this value be updated if the exposure time is higher than $\sim$ 30 minutes.
Default value \NDIT=1. 

\item \RPOW : resolving power $\lambda/\Delta\lambda$, default value is 
\RPOW=$10^5$ for HR mode and \RPOW=$10^4$ for the MR mode.

\item \EFF : total throughput of the instrument. This parameters includes the 
efficiency of the complete path (from the telescope to the detector)
and the quantum efficiency of the detector. Default value is \EFF=0.15

\item \DAPE : sky-projected angular diameter of the spectrometer aperture.
Default value \DAPE=0.8 arcsec for the HR mode and \DAPE=1.0 arcsec for 
the MR mode.
Note: the program assumes a circular aperture. Different geometries can
be accomodated by inserting the equivalent diameter of the aperture. The
result is independent on the shape of the aperture.

\item \SKYBCK : sky background at the selected wavelengths, in magnitudes AB
per square arcsec. Default values are interpolated from the following
table, which lists the approximate continuum background in between airglow
sky-lines under dark conditions.
\begin{center}
\begin{tabular}{|c|c|}
\hline
 \WL $(\mu m)$ & \SKYBCK (AB mag/arcsec$^2$)\\
 \hline
 0.36 & 22.5 \\
 0.44 & 22.5 \\
 0.55 & 21.8 \\
 0.64 & 21.5 \\
 0.80 & 20.5 \\
 1.05 & 20.0 \\
 1.25 & 19.0 \\
 1.65 & 19.0 \\
 2.20 & 19.0 \\
 \hline
\end{tabular}
\end{center}

\item \TBCK : ambient temperature, used to calculate the thermal
background, which dominates over the sky background at the longer wavelengths.
Default value \TBCK=283 K.

\item \EBCK : total emissivity of telescope and instrument, 
used to calculate the thermal
background, which dominates over the sky background at the longer wavelengths.
It includes also the atmospheric absorption.
Default value \EBCK=0.10. 

\item \DTEL : diameter of telescope, default value = 39 meters. 

\item \COBS : fractional diameter of central obscuration of the telescope,
 default value = 0.30 (i.e. 9\% in area).

\item \DPIX : physical size a detector pixel, default value = 15 $\mu$m.

\item \FXCAM : focal aperture of the camera along dispersion, default value
\FXCAM= 1.8. 

\item \FYCAM : focal aperture of the camera in the cross-dispersion 
direction, default value \FYCAM= 1.1.

\item \RON : read-out noise of the detector. Default
value \RON= 5 e$^-$/pix. 

\item \DARKCUR : dark current of the detector. Default
\DARKCUR= 18 e$^-$/pix.

\item \SLE : slit efficiency, i.e. fraction of the light from the astronomical
target falling inside the spectrometer slit. This parameter depends on
the angular size of the object, on the sky-projected
size of the slit and on the point spread function (PSF) delivered by
the telescope. The PSF depends on the seeing conditions and, if an
adaptive optics system is included, by
the performances of the adaptive optics system itself.
In this first version of the program we leave \SLE\ as a free parameter
to be manually adjusted by the user. The default value is \SLE=0.5

\end{itemize}

\ \\
The derived quantities are computed as follows.
\newcommand{\XANPIX}{\sl XANPIX}
\newcommand{\YANPIX}{\sl YANPIX}
\newcommand{\FBCK}{\sl FBCK}
\newcommand{\NBCK}{\sl NBCK}
\newcommand{\NOBJ}{\sl NOBJ}
\newcommand{\ATEL}{\sl ATEL}
\newcommand{\PIXAPE}{\sl PIXAPE}
\newcommand{\NOISEDET}{\sl NOISEDET}
\newcommand{\NOISEBCK}{\sl NOISEBCK}
\newcommand{\NOISETOT}{\sl NOISETOT}
\newcommand{\MAGLIM}{\sl MAGLIM}

\begin{itemize}

\item Telescope area
  $$ \ATEL = \pi/4 \cdot 10^4 \cdot \DTEL^2 \cdot (1-\COBS^2) 
   \ \ \ {\rm cm^2} $$

\item Equivalent pixel size, sky-projected angles
   $$ \XANPIX = 0.044 
            \left(\FXCAM\over 1.8\right)^{-1}
            \left(\DPIX\over {\rm 15 \mu m}\right)
            \left(\DTEL\over {\rm 39 m}\right)^{-1}
    \ \ \ {\rm arcsec} $$
   $$ \YANPIX = 0.072 
            \left(\FYCAM\over 1.1\right)^{-1}
            \left(\DPIX\over {\rm 15 \mu m}\right)
            \left(\DTEL\over {\rm 39 m}\right)^{-1}
    \ \ \ {\rm arcsec} $$

\item Number of pixels corresponding to spectrometer aperture
   $$ \PIXAPE = {\pi/4\cdot \DAPE^2 \over \XANPIX \cdot \YANPIX} 
     \ \ \ \ {\rm pixels} $$
Using default values of parameters yields $\PIXAPE$=248 and $\PIXAPE$=159 in
the MR and HR modes.

\item Detector noise over detector area corresponding to spectrometer aperture
  $$ \NOISEDET =
    \sqrt{\PIXAPE\cdot(\NDIT\cdot\RON^2+\DARKCUR\cdot\EXPTIME)} 
    \ \ \ \ \ e^- $$

\item Background flux in spectrometer aperture
   $$  \sigma_{sky} = {10^{(16.85-\SKYBCK)/2.5}\over \RPOW} \ \ 
     {\rm photons\ cm^{-2}\ s^{-1}\ arcsec^{-2}} $$
   $$  \sigma_{th} = {1.4\, 10^{12} \cdot \EBCK \cdot
     \exp{[-14388/(\WL\cdot\TBCK)]}
      \over \WL^3 \cdot \RPOW }  \ \
     {\rm photons\ cm^{-2}\ s^{-1}\ arcsec^{-2}} $$
\ \\
   $$  \NBCK = \EFF \cdot \ATEL \cdot \pi/4 \cdot \DAPE^2 \cdot 
     ( \sigma_{sky} + \sigma_{th})
     \ \ \ \ \  \ \ \ {\rm e^-\ s^{-1} } $$

\item Background noise in spectrometer aperture and resolution element
  $$ \NOISEBCK =\sqrt{\NBCK\cdot\EXPTIME\cdot3600} 
    \ \ \ \ \ e^- $$

\item Total noise per resolution element
   $$ \NOISETOT = \sqrt{\NOISEBCK^2 + \NOISEDET^2 + \NOBJ} 
    \ \ \ \ \ e^- $$
where $\NOBJ\ $ in the total number of photo-electrons per resolution element
produced by the object in the give exposure time.

\item Signal to noise per resolution element
   $$ \SN = {\NOBJ \over \NOISETOT} $$
which can be rewritten as
   $$ \SN^2 = {\NOBJ^2 \over\NOISEBCK^2 + \NOISEDET^2 + \NOBJ} $$
Solving for $\NOBJ\ $ yields
   $$ \NOBJ = {\SN^2 \over 2} \cdot
      \left( 1+\sqrt{1 + 4 \; {(\NOISEBCK^2 + \NOISEDET^2)\over \SN^2} }\right)
     $$

\item Object signal per resolution element on the detector
   $$  \NOBJ = {\SLE \cdot \EFF \cdot \ATEL \cdot 3600\cdot EXPTIME 
        \over \RPOW} \cdot 10^{(16.85-\MAGLIM)/2.5} \ \ 
     \ \ \ \ \  \ \ \ {\rm e^- } $$
   where $\MAGLIM\ $ is the limiting AB magnitude (i.e. the output parameter)
   computed by the program. 
\end{itemize}

Combining the last two equations
yields the following explicit expression for $\MAGLIM$

\begin{eqnarray*}
\MAGLIM = &16.85&- \ 2.5\cdot \log_{10}{\left(\RPOW\cdot \SN^2 \over 
2 \cdot 3600\cdot \EXPTIME\cdot\SLE\cdot\EFF\cdot\ATEL\right)}- \\
&+&2.5\cdot \log_{10}{\left(1+\sqrt{1 + 4 \; {\NOISEBCK^2 + \NOISEDET^2\over \SN^2} }\right)}
\end{eqnarray*}

\section{\bf Trade-off seeing-limited vs. AO}
We have used this ETC to estimate the limiting magnitudes for a compact source and 
an extended object in three different cases: seeing-limited (baseline) and two 
representative AO cases. 
For what concerns AO, we have considered a ground layer adaptive optics (GLAO, seeing improved) 
and a laser tomography adaptive optics (LTAO).
For the AO we have envisaged smaller fibers than for the baseline case.
In particular, we used $DAPE=0.8''$ for the HR seeing-limited case, $DAPE=1.0''$ for the MR 
seeing-limited, $DAPE=0.2''$ for the GLAO and $DAPE=0.1''$ for the LTAO case. 
For any instrumental configuration, we derived the limiting magnitudes both in MR and in HR mode at different S/N
levels, in various wavelengths, assuming an efficiency of 0.15 and adopting different values of 
slit efficiency according 
to the tables developed by ESO for the E-ELT exposure time 
calculator.\footnote{https://www.eso.org/observing/etc/doc/elt/etc\_spec\_model.pdf}\\
%Note that when working with fibers it is difficult to attain a low emissivity. We assumed $EBCK=0.1$, but 
%higher values of this parameter involve a decrease in the limiting magnitudes.
%In particular, an increase of emissivity of a factor of 2.5 causes a decrease of the liming magnitude in the K band
%of $\sim$ 0.5 mag.
The results are shown in Figures~\ref{fig1}, ~\ref{fig2}, ~\ref{fig3}, ~\ref{fig4} and discussed 
in the following sections.
\input{fig1}

\input{fig2}

\input{fig3}

\subsection{Compact source}
It is evident from the bottom panel of Figure~\ref{fig1}, that in the high S/N HR mode there is a 
strong advantage in using LTAO in the K band and  a marginal one in the H band. 
At lower S/N levels (see the bottom panel of Figure~\ref{fig2}), 
LTAO allows one to achieve higher performances at shorter wavelengths (down to the J).\\
In MR mode, LTAO sensitivity is the highest already at J in high S/N regime 
(see the bottom panel of Figure~\ref{fig3}) and again its advantage over seeing-limited and GLAO set-ups shifts to
shorter wavelengths (Y bands) 
at lower S/N levels (see the bottom panel of Figure~\ref{fig4}).\\
Note that the sensitivity in seeing-limited mode drops in the K band because of the sky background, while in the other IR bands the read out noise 
and the dark current are dominant and the binning on chip is not possible. 
\input{fig4}

\subsection{Extended source}
We have assumed that the encircled energy (fraction of target within the aperture diameter) for a typical high redshift galaxy is 
100$\%$, 50$\%$ and 25$\%$ in the seeing-limited, GLAO and LTAO cases, respectively.
From the top panels of  Figures~\ref{fig1}, ~\ref{fig2}, ~\ref{fig3}, ~\ref{fig4} it is evident that there is no advantage 
in using AO. Although there is a marginally better sensitivity in the K band with LTAO (low S/N in the HR mode and any 
S/N level in the MR mode), it is important to remark that using two different observing set-ups (seeing-limited
and LTAO) for an extended object implies that different fractions of the same object are observed, due to the different 
aperture diameters used.

\section{Conclusions}
It is important to remark that the dimension of the fibers changes in different observing modes,
becoming smaller for the AO modes than for the baseline case, and the aperture diameters play an important role.\\
The simulations based on the ETC demonstrate that at visual/blue wavelengths HIRES will be 
better designed as a seeing-limited instrument.
In fact, in all the analyzed cases (compact or extended source, resolution power at any S/N level) 
AO does not allow any performance enhancement at these wavelengths.\\
There is almost no advantage in using a seeing improved instrument (GLAO) 
in comparison with being seeing-limited.\\
On the other hand, in the infrared region the LTAO can yield a better sensitivity. 
In fact, it is the most efficient instrumental configuration for compact sources. 
The critical wavelength, where LTAO starts to out-perform seeing-limited observations,
progressively shifts from the K to the Y band for decreasing resolving powers and/or 
signal-to-noise.\\
The performance for extended objects is different.
Even if in some cases (low S/N levels and MR) 
there is a marginal advantage using LTAO in the K band, probably it is not enough to compensate the 
disadvantage due to observing different fractions of the object, as mentioned in the previous section.\\
The only way to optimize the performances for all type of observations is to foresee both seeing-limited 
and AO-fed observing modes in the infrared arms of HIRES. 

\section{Acknowledgments}
This work was financially supported by INAF through the grants 
"TECNO-2011" and "TREX-2011"

\input{biblio}
\end{document}

%% file: fig1.tex
\begin{figure}[ht]
\begin{center}
\includegraphics[scale=0.7]
 {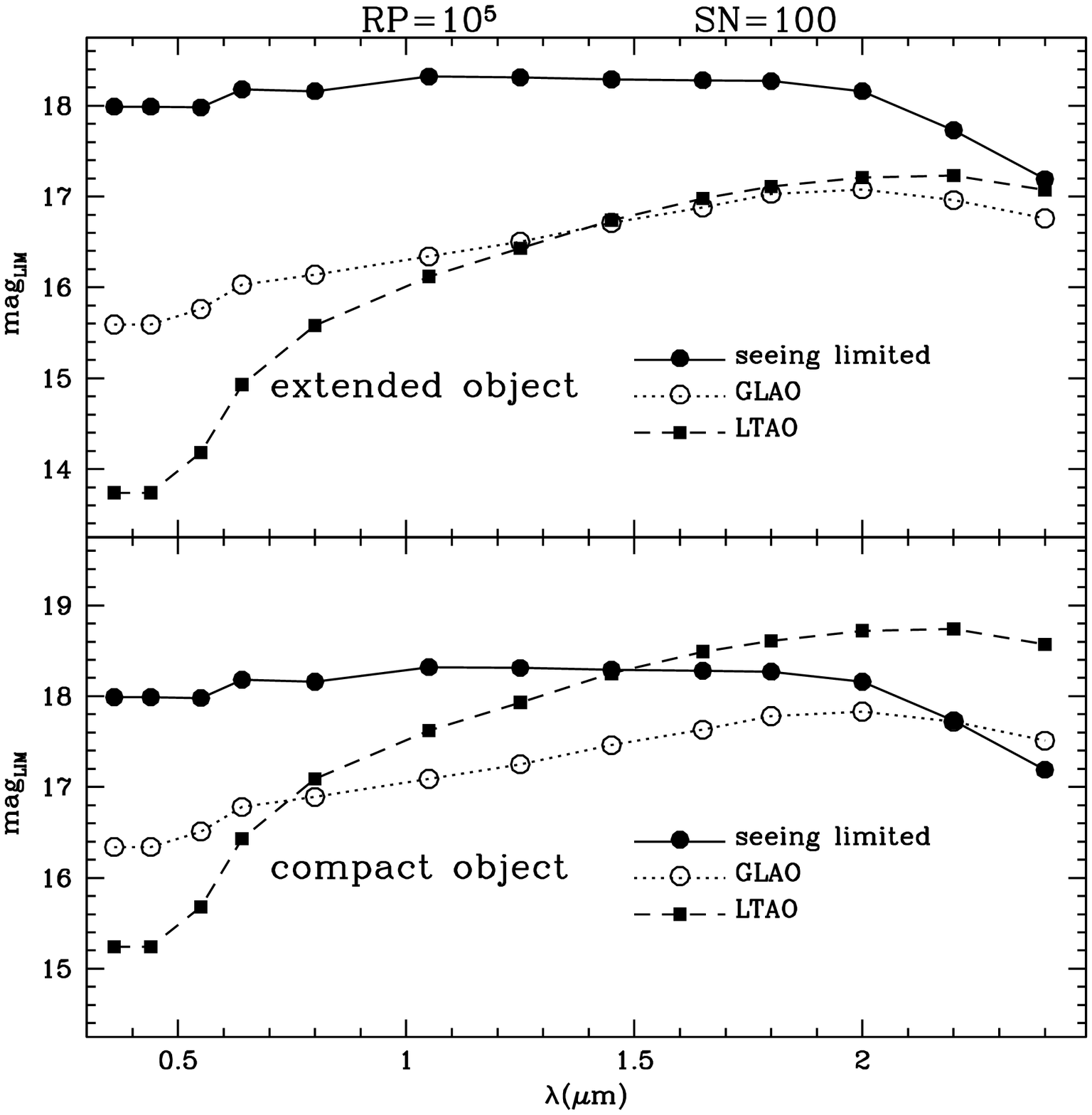}
\end{center}
\vspace{-1cm}
\caption[]{Limiting magnitudes in AB units for a compact (bottom) and an 
extended (top) source at $RP=10^5$ and $S/N=100$ in the seeing limited (filled
circles, continuos line), GLAO (open circles, pointed line) and LTAO (filled 
squares, dashed line) cases. Note the key role played by the fiber size (different
aperture diameters for different observing set-ups).
}
\label{fig1}
\end{figure}

%% file: fig2.tex
\begin{figure}[ht]
\begin{center}
\includegraphics[scale=0.7]
 {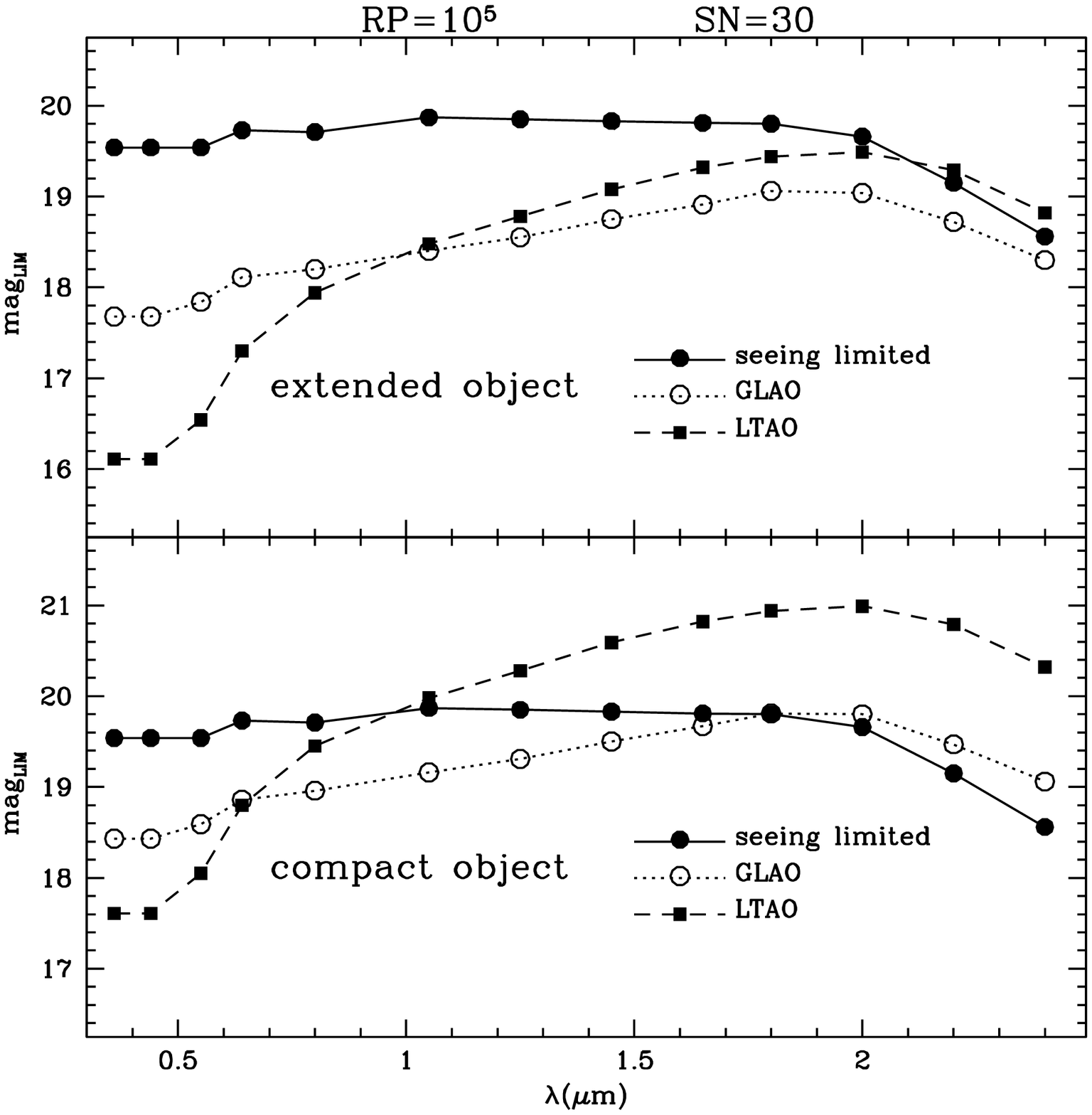}
\end{center}
\vspace{-1cm}
\caption[]{Limiting magnitudes in AB units for a compact (bottom) and an 
extended (top) source at $RP=10^5$ and $S/N=30$ in the seeing limited (filled
circles, continuos line), GLAO (open circles, pointed line) and LTAO (filled 
squares, dashed line) cases. Note the key role played by the fiber size (different
aperture diameters for different observing set-ups).
}
\label{fig2}
\end{figure}

%% file: fig3.tex
\begin{figure}[ht]
\begin{center}
\includegraphics[scale=0.7]
 {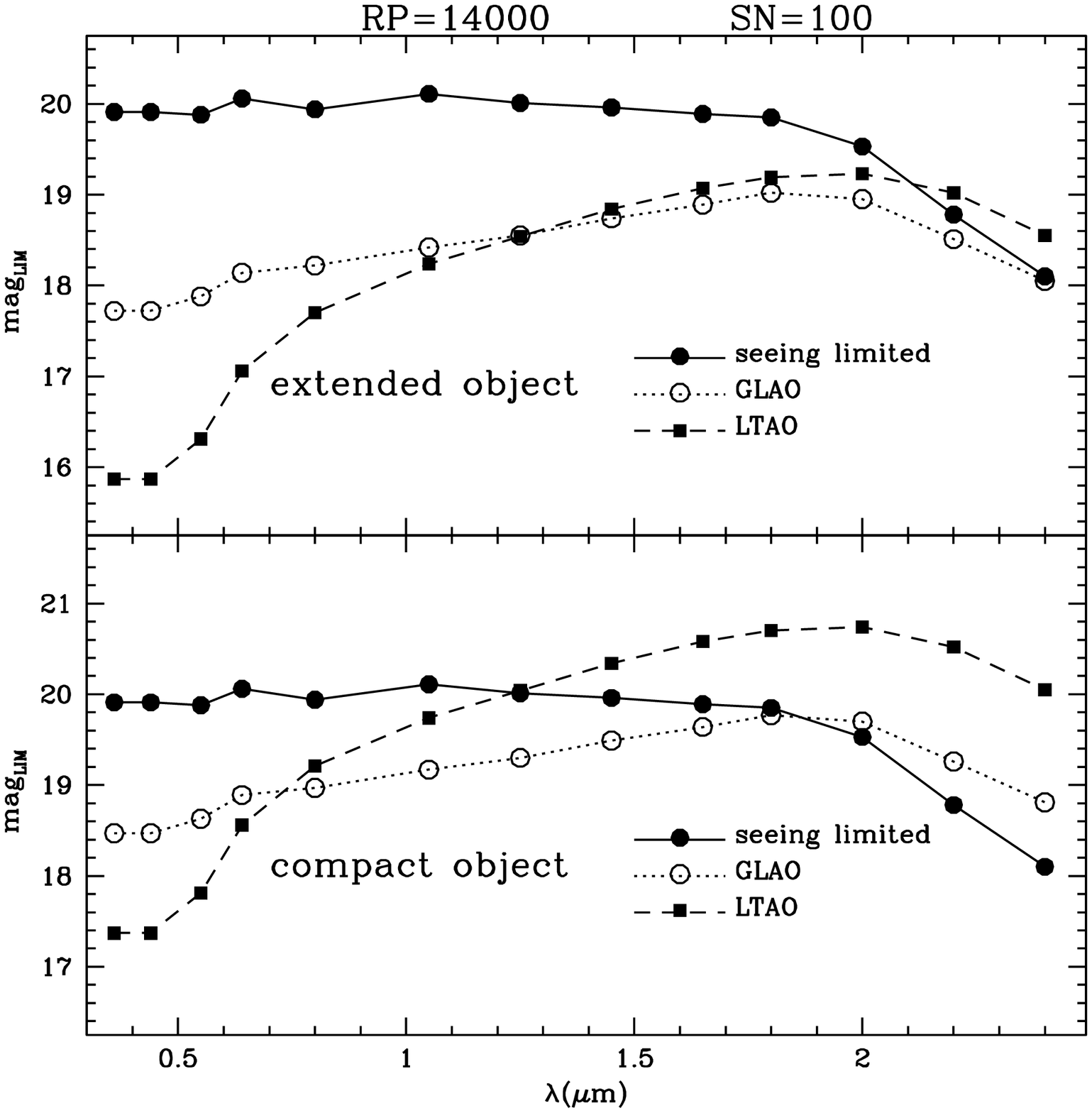}
\end{center}
\vspace{-1cm}
\caption[]{Limiting magnitudes in AB units for a compact (bottom) and an 
extended (top) source at $RP=14000$ and $S/N=100$ in the seeing limited (filled
circles, continuos line), GLAO (open circles, pointed line) and LTAO (filled 
squares, dashed line) cases. Note the key role played by the fiber size (different
aperture diameters for different observing set-ups).
}
\label{fig3}
\end{figure}

%% file: fig4.tex
\begin{figure}[ht]
\begin{center}
\includegraphics[scale=0.7]
 {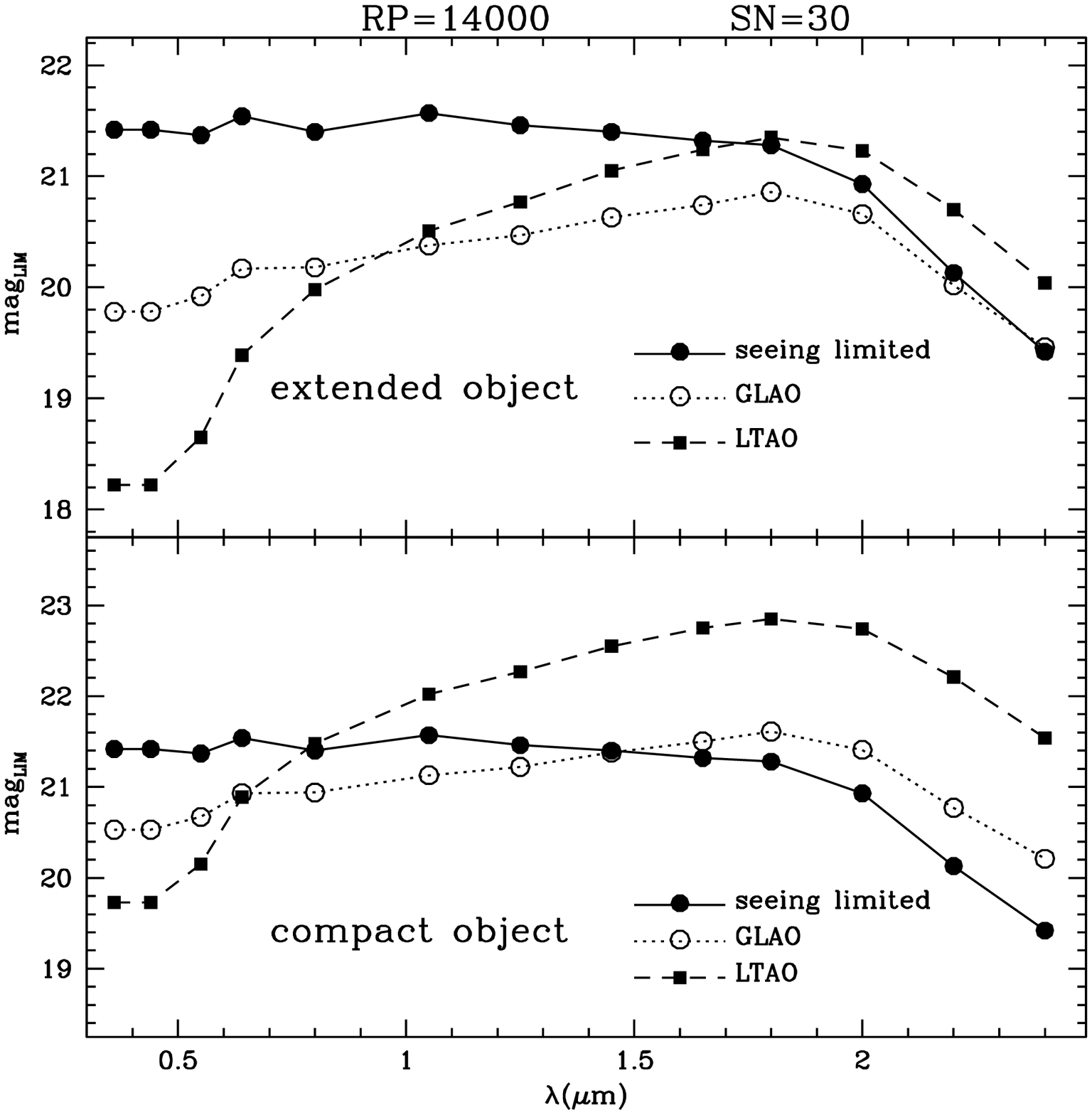}
\end{center}
\vspace{-1cm}
\caption[]{Limiting magnitudes in AB units for a compact (bottom) and an 
extended (top) source at $RP=14000$ and $S/N=30$ in the seeing limited (filled
circles, continuos line), GLAO (open circles, pointed line) and LTAO (filled 
squares, dashed line) cases. Note the key role played by the fiber size (different
aperture diameters for different observing set-ups).
}
\label{fig4}
\end{figure}